# Non-Equilibrium Molecular Dynamics Study of

# Thermal Energy Transport in Au-SAM-Au junctions

**Tengfei Luo**

2555 Engineering Building

Mechanical Engineering

Michigan State University

East Lansing, MI 48824

Email: luotengf@msu.edu

**John R. Lloyd**

ASFC 1.316.C RRMC

Rapid Response Manufacturing Center

The University of Texas Pan American

Edinburg, TX 78539

Email: lloyd@egr.msu.edu

**Abstract:**

Non-equilibrium molecular dynamics (NEMD) simulations were performed on Au-SAM (self-assembly monolayer)-Au junctions to study the thermal energy transport across the junctions. Thermal conductance of the Au-SAM interfaces was calculated. Temperature effects, simulated external pressure effects, SAM molecule coverage effects and Au-SAM bond strength effects on the interfacial thermal conductance were studied. It was found that the interfacial thermal conductance increased with temperature increase at temperatures lower than 250K, but it did not have large changes at temperatures from 250K to 400K. Such a trend was found to be similar to experimental observations on similar junctions. The simulated external pressure did not affect the interfacial thermal conductance. SAM molecule coverage and Au-SAM bond strength were found to significantly affect on the thermal conductance. The vibration densities of state (VDOS) were calculated to explore the mechanism of thermal energy transport. Interfacial thermal resistance was found mainly due to the limited population of low-frequency vibration modes of the SAM molecule. Ballistic energy transport inside the SAM molecules was confirmed, and the anharmonicity played an important role in energy transport across the junctions. A heat pulse was imposed on the junction substrate, and heat dissipation inside the junction was studied. Analysis of the junction response to the heat pulse showed that the Au-SAM interfacial thermal resistance was much larger than the Au substrate and SAM resistances separately. This work showed that both the Au substrate and SAM molecules transported thermal energy efficiently, and it was the Au-SAM interfaces that dominated the thermal energy transport across the Au-SAM-Au junctions.

## 1. Introduction

Molecular electronic devices have become more and more popular nowadays, and among them, the SAM-metal and SAM-semiconductor junctions have drawn much attention. There has been considerable interest focused on the electronic and structural properties of SAM-solid junctions [1-4]. However, the studies of thermal properties of such junctions are limited, and knowledge of thermal transport in these junctions is critical to the growing fields of molecular electronics and small molecule organic thin film transistors. Moreover, for thermal energy transport in systems at nanoscale, the interface becomes more important than its macroscopic counterpart [14]. Ge et. al [5] measured the transport of thermally excited vibrational energy across planar interfaces between water and solids that have been chemically functionalized with SAM using the time-domain thermoreflectance. Wang et. al [6] studied heat transport through SAM of long-chain hydrocarbon molecules anchored to a gold substrate by ultrafast heating of the gold. Patel et. al [7] studied interfacial thermal resistance of water-surfactant-hexane systems using NEMD. Wang et. al [35] measured thermal conductance of Au-SAM-GaAs junctions using the $3\omega$ technique. Segal et. al. [41] studied the thermal transport through alkane molecules using a quantum mechanical approach. For metal-nonmetal interfaces, the experimental thermal conductance is reported to be $8 < G < 700 \ MW/(m^2K)$ [37-39].

SAM's are usually grown on metals [8,9] or semiconductors [10] forming solid-molecule junctions. In this work, thermal transport in Au-SAM-Au junctions with alkanedithiols being the SAM molecules is studied using molecular dynamics (MD) simulations. SAM-Au junctions are chosen because their structural properties, including the absorption site, tilt angle, coverage and etc., are well documented [8,9,11,12], and a set of reliable classical potentials for MD simulation is available [13] which facilitate MD studies of thermal transport in such junctions.

In this work, where the metal-nonmetal junctions exist, electron transport is largely depressed by the nonmetal material. As a result, this work focuses on the phonon (quanta of the lattice vibrational field [15]) part of thermal transport and calculates the phonon thermal conductance of the Au-SAM interfaces. For phonon thermal conductance calculations, classical MD with appropriate potential functions has been demonstrated to be a powerful method [16-20]. However, to our knowledge, except our previous work [21] which used equilibrium MD (EMD) and Green-Kubo method to calculate thermal conductivities of Au-SAM-Au junctions, no MD simulation has been done to investigate the thermal transport in the metal-SAM-metal junctions. In this work, NEMD simulations are performed on Au-SAM-Au junctions and the Au-SAM interfacial thermal conductance is calculated. The thermal conductance is calculated versus different simulation cell sizes, system mean temperatures, simulated normal pressures, SAM molecule coverage and Au-SAM bond strengths. The calculated data and their trends are compared with available experimental measurements. VDOS's are studied to explore the mechanism of the thermal energy transport in the junctions. Heat dissipation inside a junction which subjected to a heat pulse on the substrate is also simulated.

## 2. Theory and Simulation

Classical MD is a computational method that simulates the behavior of a group of atoms by simultaneously solving Newton's second law of motion (eq.(1)) for the atoms with a given set of potentials.

$$-\nabla\phi = \vec{F} = m\frac{d^2\vec{r}}{dt^2}.\qquad(1)$$

In an NEMD simulation for thermal conductance calculation, either a constant temperature difference [22-25] or an applied heat flux [7,25-29] is imposed by altering the atomic dynamics in localized heat sink and source regions. In the constant temperature difference method, the temperatures of the sink and source regions are controlled at the different values by scaling the atomic velocities in these two regions. The energy differences before and after the velocity scaling processes in both the sink and source regions generate heat fluxes. Although the velocity scaling is an abrupt disturbance to the simulation system, it has proven to be a valid algorithm which does not significantly modify the local thermal equilibrium of the sink and source region [22,24,36]. In the heat flux method of this work, the velocities of the atoms in the sink and source regions are scaled so that the same amount of energy is added to the sources and taken out from the sinks. The amount of energy difference before and after the scaling is determined by the energy difference between the fastest atom in the sink region and the slowest atom in the source region. In this work, the second algorithm is mainly used while the first one is also used for comparison purpose.

In an NEMD simulation, after the simulated system reaches a steady state, the Au-SAM interfacial thermal conductance can be calculated by Fourier's law (eq. 2).

$$J = G\Delta T \tag{2}$$

where $\Delta T$ is the temperature difference at the interface, $J$ is the heat flux which is defined as the amount of thermal energy transfer rate per unit area perpendicular to the direction of the heat flux and $G$ is the thermal conductance in the heat flux direction.

A typical set-up of the simulation system of this work is shown in Figure 1. The simulation system consists of three Au substrates with SAM connecting them. The SAM is made of octanedithiols ($-S-(CH_2)_8-S-$). The procedure of preparing the junction was same as that described in ref. [21]. Periodic boundary conditions (PBC) are used in all three spatial (x-, y-, z-) directions. With PBC in z-direction, the left substrate and the right substrate make up a united substrate which is exactly the same as the substrate in the center. Four Au layers at the center are chosen as the sink region and two Au layers at each end are chosen as the source region. Atoms in these two regions are subject to velocity scaling while all the rest of the atoms move freely.

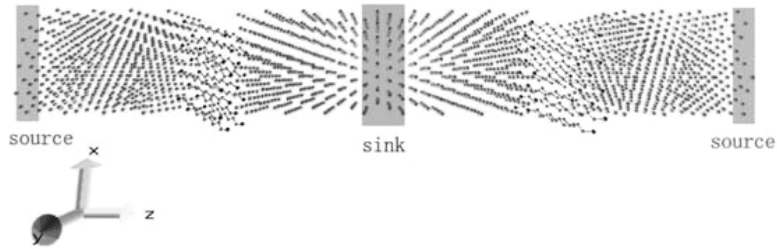

Figure 1. A typical set-up of the simulation system.

For the system shown in Figure 1, eq. (2) can be expressed as

$$J = \frac{dE}{2A \bullet \Delta t} = G\Delta T \tag{3}$$

where $\Delta t$ is the time interval between two consecutive velocity scalings, $A$ is the area of the system perpendicular to the heat flux and $dE$ is the energy taken out from the heat sink and put in to the heat source. Thermal conductance errors are estimated by the error propagation relation and the relative errors are calculated to be 25-30%.

In this work, a system which contains 4 molecules in each SAM and 12 gold atoms in each substrate layer is defined as a unit cell (please note that only the dimensions in the x- and the y-directions of the unit cell are defined while the z-direction dimension not defined). It has a dimension of 9.99 $\overset{\circ}{A}$ in the x-direction and 8.652 $\overset{\circ}{A}$ in the y-direction. The simulation supercells are obtained by expanding the unit cell in x and y directions. The system in Figure 1 is a 2x2 supercell.

In this work, the well established Hautman-Klein model [13] is employed. In this model, the light hydrogen atoms of the hydrocarbon molecules are not simulated explicitly but incorporated into the carbon backbones. Bond stretching, bond bending and Ryckaert-Bellemans torsion potentials are used in the SAM molecules for the bonded interactions. Morse potentials are used to simulate the interactions between S atoms and Au substrates [30,31] and the interactions among Au atoms [32]. The Lennard-Jones (LJ) potentials together with the Lorentz-Berthelot mixing rule [33], $\varepsilon_{ab} = \sqrt{\varepsilon_a \varepsilon_b}$ , $\sigma_{ab} = \frac{1}{2}(\sigma_a + \sigma_b)$ , are used to simulate long distance van de Waals interactions.

Detailed formulations and parameters of these potentials can be found in ref. [21].

The simulation procedure is: (1) All atoms start moving from their equilibrium positions with random initial velocities. (2) Thermostats are applied to the whole system to make sure that the system reaches the target mean temperature. (3) Thermostats are then released, and an equilibration period is performed. (4) Velocity scalings are applied to the heat sink and source to create thermal non-equilibrium state. (5) After the system reaches a steady state, a production period during which the heat current and temperatures at different z coordinates are recorded. (6) Thermal conductance was calculated according to eq. 3. For all the simulations, neighborlists were used to speed up the calculations, and the time step was set to 0.5 fs. For each case, five runs with different initial conditions are performed, and the thermal conductance values are averaged over the results from the five runs. In each run, the interfacial temperature difference, $\Delta T$ , are averaged over the four Au-SAM interfacial temperature differences.

## 3. Results
### 3.1 Finite Size Effect

To investigate the finite size effect in all three directions, interfacial thermal conductance values of systems with different cross section sizes and different substrate thicknesses are calculated. Table 1 shows the results from calculations using the constant temperature difference method, and Table 2 shows results from calculations with the heat flux method.

| Case | Source temp. (K) | Sink temp. (K) | Cross section size | Cross section area ($\overset{\circ}{A} \times \overset{\circ}{A}$) | Junction length ($\overset{\circ}{A}$) | $G$ ($MW/(m^2 K)$) |
|------|------|------|------|------|------|------|
| 1 | 130 | 70 | 2x2 | $19.98 \times 17.30$ | 136.20 | 340 |
| 2 | 130 | 70 | 2x2 | $19.98 \times 17.30$ | 251.34 | 352 |
| 3 | 130 | 70 | 3x3 | $29.97 \times 25.95$ | 136.20 | 344 |
| 4 | 140 | 60 | 2x2 | $19.98 \times 17.30$ | 136.20 | 330 |
| 5 | 150 | 50 | 2x2 | $19.98 \times 17.30$ | 136.20 | 342 |

Table 1. Thermal Conductance of Systems with the Constant Temperatures Method

| Case | Cross section size | Cross section area ($\overset{\circ}{A} \times \overset{\circ}{A}$) | Junction length ($\overset{\circ}{A}$) | $G$ ($MW/(m^2K)$) |
|------|------|------|------|------|
| 6 | 2x2 | $19.98 \times 17.30$ | 136.20 | $348 \pm 80$ |
| 7 | 2x2 | $19.98 \times 17.30$ | 251.34 | $360 \pm 85$ |
| 8 | 3x3 | $29.97 \times 25.95$ | 136.20 | $352 \pm 80$ |

Table 2. Thermal Conductance Systems with Heat Fluxes Imposed

In Table 1, the interfacial thermal conductance values from systems with different sizes are close to each other. For case 1 and 3, which are only different in cross section area, the interfacial thermal conductance values are almost the same. This suggests that the finite size effect is not important in the x- and y-directions for the 2x2 systems. The steady state temperature profile of case 4 is shown in Figure 2 (a). For case 1 and 2, which are only different in substrate thicknesses, the interfacial thermal conductance of the system with thicker substrate has a slightly higher value.

Since thicker substrates allow phonons with larger mean free path to be excited in the simulation, more phonons are involved in thermal energy transport in the system with thicker substrates than that with thinner substrates. However, it is also found that the thermal conductance difference between case 1 and 2 is only 3.5%. As a result, we can ignore the substrate thickness effect when the systems have lengths of $136.2 \overset{\circ}{A}$. In Figure 2 (b), which shows the steady state temperature profile of case 2, the substrate temperature profiles away from the heat sink and source or the SAM-Au interfaces are very flat. Such an observation suggests that the thermal conductance in the Au substrates is very large compared to the thermal conductance of the Au-SAM interfaces where large temperature gaps exist. The temperature profiles are nonlinear near the sink and the source regions due the artificially altered dynamics by the velocity scalings. Also, the temperature profiles become nonlinear near the interfaces which suggest that the dynamics of the Au atoms near the interfaces are different than those inside the substrate due to the surface effect. It should be noted that the temperature nonlinearities in Au substrates do not affect the validity of applying eq. 2 on the Au-SAM interface since only the temperature difference between two points are involved to calculate the interfacial thermal conductance.

Simulations with different heat sink and source temperatures were also performed (see case 1, 4 and 5), and the calculated thermal conductance values are not very different from one another. This suggests an almost linear relation between the heat flux and the interfacial temperature difference, which demonstrates the validity of the Fourier's law on the Au-SAM interface. The simulations with the heat flux method yield results (Table 2) that are similar to those from constant temperature difference method. Although these two nonequilibrium algorithms yield similar results, the large fluctuations of heat fluxes in the constant temperature difference method (Figure 3 (a)) lead to relative errors larger than 1. Moreover, it is very difficult to make sure that the simulation system reaches a state in which the mean energy taken out from the heat sink equals that put into the heat source. The heat flux method does not suffer from the above two problems (Figure 3(b)), and thus the rest of the work uses the heat flux method. As discussed above, a 2x2 system with thickness of $136.20 \overset{\circ}{A}$ is enough to ignore the finite size effect, and such systems are used throughout the rest of the simulations.

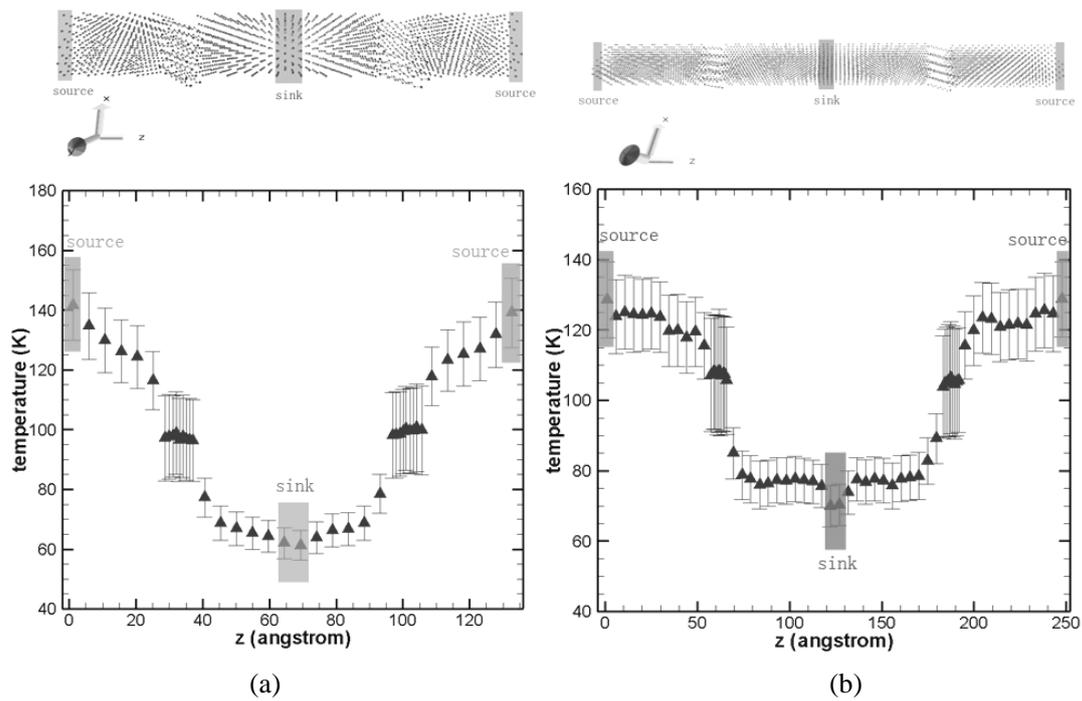

(a)            (b)

Figure 2. Steady state temperature profiles: (a). Temperature profile of a 2x2 system with a junction thickness of $136.20 \, \overset{\circ}{A}$. (b). Temperature profile of a 2x2 system with a junction thickness of $251.34 \, \overset{\circ}{A}$.

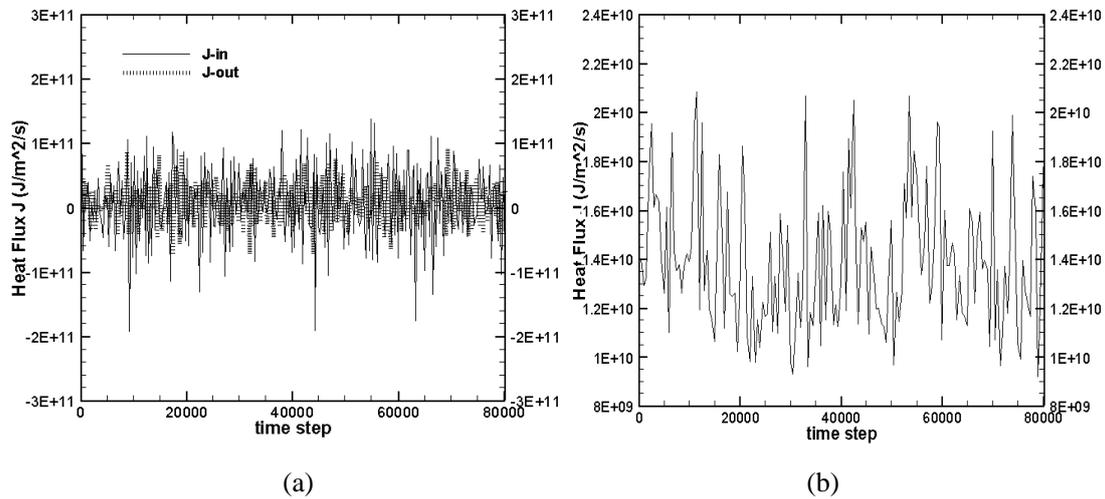

(a)            (b)

Figure 3. Heat fluxes: (a) heat flux of heat source and heat sink in a simulation with constant sink and source temperatures, (b) heat flux of a simulation with the heat flux method.

### 3.2 Temperature Effect

For NEMD simulations in which there are temperature differences, an equilibrium temperature of the system is not defined. Systems with different mean temperatures are studied. The thermal conductance values are plotted against the mean temperatures, as shown in Figure 4.

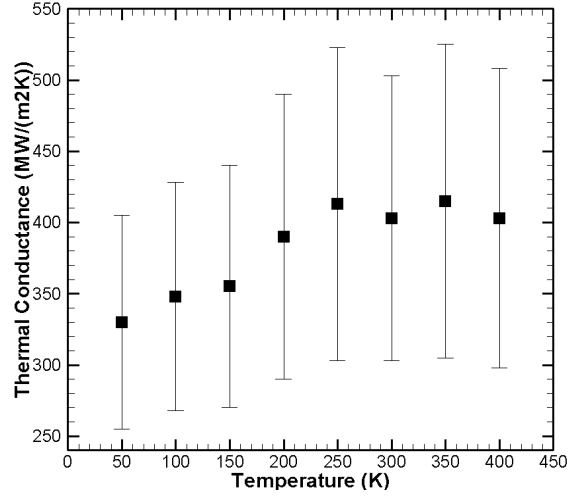

Figure 4. Au-SAM interfacial thermal conductance at different mean temperatures.

The interfacial thermal conductance increases with temperature increase when the mean temperatures are below 250K. The thermal conductance values do not have significant changes at temperatures above 250K. Such a trend of thermal conductance is similar to that found in the experiments by Wang et. al. [35] on GaAs-SAM-Au junctions and from our previous EMD study of the same Au-SAM-Au junctions [21]. The plateau in Figure 4 appears at a higher temperature than those in the aforementioned two references. It should be noted that in the nonequilibrium steady state, the two inner Au-SAM interfaces are at temperatures lower than the system mean temperature, while the outer two interfaces are at temperatures higher than the system mean temperature (see Figure 2). As a result, we estimate that the thermal conductance stops increasing with temperature at a temperature around 200 K, which is close to that found in the EMD study [21]. It should also be noted that the GaAs-SAM junctions and the Au-SAM junctions have different degrees of anharmonicity. Since the anharmonicity is a factor that will counter affect thermal transport across the junction (will be discussed in the next paragraph), it is easy to understand why the plateau appears at different temperatures in different junctions. In the experiment on Au-SAM junctions, which was conducted by Wang et. al [6], an estimated interfacial thermal conductance of $220 \pm 100 MW/(m^2 K)$ at $800^{\circ}C$ was reported. Considering the approximations made in estimating the thermal conductance in ref. [6], our high temperature data ($400 \pm 120 MW/(m^2 K)$) are in reasonable agreement with their estimated value. If we assume that the thermal conductance of the Au-SAM-Au junction is dominated by the Au-SAM interfaces, the junction thermal conductance in this work is approximately $200 \pm 60 MW/(m^2 K)$ at high temperatures. In ref. [41] by Segal et. al., where a quantum mechanical model was used to predict the thermal conductance of an alkane chain attached to two electrodes, a thermal conductance of about $3.5 \times 10^{-11} W/K$ per alkane chain at 300K was reported. Considering the area per chain to be $2.16 \times 10^{-19} m^2$ in our simulations, the thermal conductance per chain of our study is about $4.3 \times 10^{-11} W/K$, which is in good agreement with the prediction by Segal et. al [41].

To further analyze the thermal energy transport in the Au-SAM-Au junctions, VDOS's are calculated by performing Fourier transform of the velocity autocorrelation functions (VAF) [43] according to eq. 4.

$$D_\alpha(\omega) = \int_0^\tau \Gamma_\alpha(t) \cos(\omega t) dt \qquad (4)$$

where $\omega$ is frequency, $\alpha$ refers to different elements, $D_\alpha(\omega)$ is the VDOS at frequency $\omega$, and $\Gamma_\alpha(t)$, which is give in eq. 5, is the VAF of element $\alpha$.

$$\Gamma_\alpha(t) = \langle v(t) \bullet v(0) \rangle \qquad (5)$$

The VDOS's of surface Au (the top layer of Au substrate at the Au-SAM surface), S, C1(the 1st carbon atom from the sulfur head), C2 (the 2nd carbon atom from the sulfur head) and C4 (the 4th carbon atom from the sulfur head) atoms at temperatures of 50K, 150K, 250K and 350K. The VAF's are obtained from equilibrium runs at the aforementioned temperatures.

The results are presented in the Figure 5. The VDOS's are not only proportional to the population of vibration modes but also proportional to temperature since they are proportional to the square of the velocity [42]. To make the VDOS's of different elements comparable, VDOS's of different elements are weighted by their respective mass.

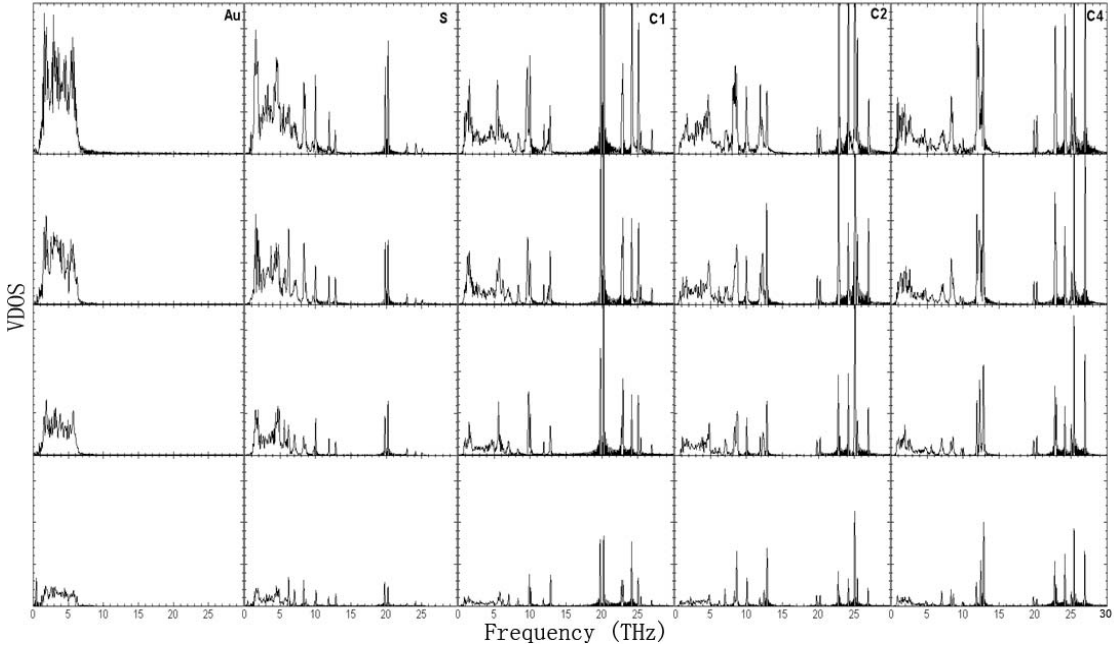

Figure 5. VDOS's of surface Au (1st column), S (2nd column), C1(3rd column), C2(4th column) and C4(5th column) atoms at temperatures of 350K (top row), 250K (2nd row), 150K (3rd row) and 50K (bottom row). C1 refers to the 1st carbon atom from the sulfur head, C2 refers to the 2nd carbon atom from the sulfur head and C4 refers to the 4th carbon atom from the sulfur head.

It can be seen that the populations of vibration modes at all frequencies of all elements grow with temperature increase. As a result, more phonons are involved in thermal energy transport, which leads to the increase of thermal conductance with temperature. However, this does not explain why the thermal conductance stops increasing at high temperatures. To facilitate analysis, we divide the molecular (including S and C atoms) vibration modes into three regions: low-frequency (LF) modes ranging from 0 to 15THz, intermediate-frequency (IF) modes ranging from 15 to 30THz and high-frequency (HF) modes with frequencies higher than 30THz. It is found that the HF modes are not excited at temperature up to 400K, so only the LF and IF modes are visualized. In Figure 5, both the substrate Au atoms and the SAM molecule atoms have largely populated LF modes, leading to a resonance type of thermal transport between the Au substrates and SAM molecules. It also can be seen that the LF modes extend overall all the molecular elements, suggesting a highly delocalized feature of these modes. The IF modes could also transport thermal energy, however, the direct coupling between the IF modes to the Au substrate is not possible since there are no vibration modes in the Au substrate with frequencies ranging from 15 to 30THz. However, the anharmonicity makes the LF and IF phonon-phonon interactions possible, and thus felicitates energy communication between LF modes and the IF modes.

One can picture the following thermal energy transport mechanism: some energy of the LF modes is transferred to IF modes at one interface due to phonon-phonon interactions. Then IF modes carry the energy through the SAM molecules and release the energy to LF modes at the other interface. Such a tunneling-like transport decreases exponentially with chain-length increase [41]. However, our previous EMD simulations on Au-SAM-Au found that the thermal conductance did not depend on molecule chain-length [21], which suggests that the tunneling-like transport does not play an important role in the thermal transport through the Au-SAM-Au junctions with SAM molecular chain-length of 8 or more carbon atoms. To increase the thermal conductance of the solid-SAM junctions, using substrate materials which can utilize the IF vibration modes for resonance type transport should be an effective strategy.

In this work, the potentials used to describe the SAM intramolecular interactions are mainly harmonic interactions [21]. Although the LJ and Ryckaert-Bellemans Torsion potential could result in anharmonicity, the effect is believed to be small because these interactions are very weak. Segal et. al [41] compared the full alkane force field (with anharmonicity) and the pure harmonic model, and it was found that the limited anharmonicity does not affect the thermal transport along the alkane chain. So, we believe thermal energy transport inside the SAM molecule is dominantly harmonic (ballistic transport), which transports thermal energy efficiently. This explains why the temperature difference inside the SAM molecule is very small (see Figure 2). However, the Morse bonds, which are used to describe the Au-Au and Au-S interactions, are not harmonic, resulting in anharmonicity. The anharmonicity which scatters phonons counter affects the thermal energy transport from Au into SAM molecule [40]. It should also be noted that the anharmonicity increases with temperature increase, and this explains why the thermal conductance does not keep increasing at high temperatures.

From Figure 2, we can see that the largest temperature drops exist at the Au-SAM interfaces. We can also see from Figure 5 that there are strong couplings between LF molecule modes and the Au substrate. It should be noted that the ratio of the number of the Au substrate atoms to the numbers of the SAM molecule atoms is 3.6. As a result, the SAM molecular spectra work as bottlenecks, which have much smaller vibration modal populations compared to that of the Au substrate (one could imagine multiplying the Au spectra by 3.6 and examining the overlap between Au spectra and SAM molecule spectra). As a result, the populations of the SAM molecular LF modes that can resonate with the substrate vibration modes are limited, and this leads to relatively large interfacial thermal resistance compared to the resistances of Au substrates and SAM molecules.

### 3.3 Simulated Pressure Effect

The normal pressure effect on the thermal conductance is simulated in this study by decreasing the dimensions of the simulation supercells in the z-directions. In this way, the junction would feel "pressure" as it is compressed by the decreased cell sizes. Simulations were performed at mean temperatures of 100K and 300K. The calculated Au-SAM interfacial thermal conductance values are plotted in Figure 6. It is found that the thermal conductance does not depend on the simulated external pressure. From structural point of view, although the junction is under pressure, due to the flexibility of the chain-like SAM molecules, the structure can adapt itself to the small dimension changes in the z-direction so that the dynamics at the SAM-Au interface are not changed much. Thus, neither the vibration modes nor the strength of the anharmonicity should change much. As a result, the phonon transport in the junction remains the same, and this leads to the same thermal conductance.

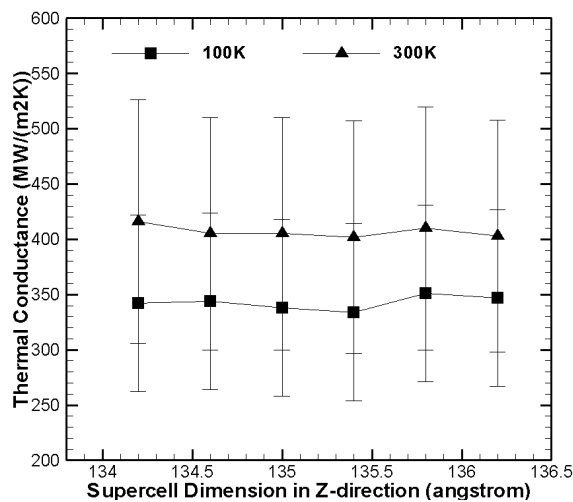

Figure 6. Interfacial thermal conductance versus simulation supercell thicknesses.

### 3.4 Coverage Effect

The SAM molecular coverage effect on thermal conductance is also studied by changing the number of alkanedithiol molecules on the substrates. With PBC in x- and y-directions, every alkanedithiol molecule is equivalent in position. Molecules are deleted symmetrically so as to keep symmetries. Au-SAM interfacial thermal conductance of the 2x2 systems with 16, 14, 12, 10, 8 alkanedithiol molecules in each SAM are calculated at a mean temperature of 100K and the results are presented in Figure 7.

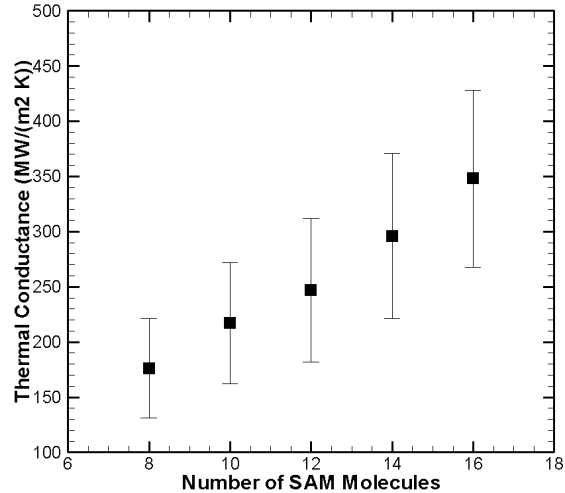

Figure 7. Coverage dependence of Au-SAM interfacial thermal conductance at 100K.

In Figure 7, it is seen that the thermal conductance increases when the number of the SAM molecules increases. In the Au-SAM-Au junctions, thermal energy is transported from one substrate to another through the SAM molecules. These molecules work as "channels" through which energy passes. When the number of molecules decreases, the number of thermal transport channels decreases. Due to the weak intermolecular interactions (van de Waals interactions), the intermolecular thermal energy transport is difficult. This makes the "channels" relatively isolated. Since each molecule has equal capacity of transporting thermal energy, the thermal conductance is almost a linear function of the number of molecules. The observations in Figure 7 can also be explained from the vibration point of view. As discussed in Section 3.2, the populations of LF molecular vibration modes work as bottlenecks which limit the thermal transport efficiency. As the number of molecules decreases, the total population of the LF molecular vibration modes to couple with Au substrate decreases, leading to the decrease of thermal conductance.

### 3.5 Au-SAM Bond Strength Effect

As discussed in Section 3.4, each SAM molecule works as an isolated thermal transport "channel". As a result, the contacts of the molecules to the Au substrate are important parameters which could influence the energy transport efficiency of each "channel". To study the effect of the Au-SAM bond strength on the thermal energy transport across the interfaces, the binding energy ($De$) of the Morse potential, which describes the Au-S bond, was increased by 10%, 20% and 30%. The calculated interfacial thermal conductance data are plotted in Figure 8. It can be seen that the interfacial thermal conductance increases as the Au-SAM binding energy increases. The VDOS's are calculated for the surface Au atoms and the S atoms from the MD run with the original bond strength ($1 \cdot De$) and that with $1.3 \cdot De$ (Figure 9). It can be seen, that the vibration modes in the Au spectra with frequencies lower than 5THz shift to the left and those with frequencies higher than 5THz shift to the right when the bond strength is increased. The S spectra also exhibit similar trends. It is easy to understand the spectra shifts to the higher frequencies due to the increased bond strength, but it is not intuitive to believe that some modes should shift to lower frequencies. The spectra shifts can potentially change the LF modes resonance between Au and S atoms, and thus change the interfacial thermal conductance. However, because the shifts are so tiny, the thermal conductance change due to the spectra shift is expected to be small. On the other hand, due to the increased bond strengths, the motions of the surface Au atoms and the S atoms are more confined around the potential minimum (equilibrium position), and thus the anharmonicity at the interface is decreased. This leads to smaller phonon scattering at the interface due to the anharmonicity, hence higher thermal conductance. As a result, we believe that the thermal conductance change due to bond strength change is mainly because of the change of the anharmonicity from the Au-S bond. Some modes shifting to lower frequencies could also be a result of the change of the anharmonic scattering at the interface.

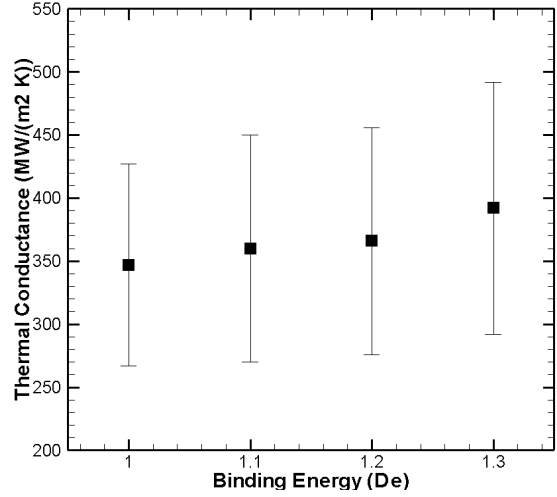

Figure 8. Binding Energy Dependence of Au-SAM Interfacial Thermal Conductance at 100K. $De$ is the binding energy of the original Morse potential.

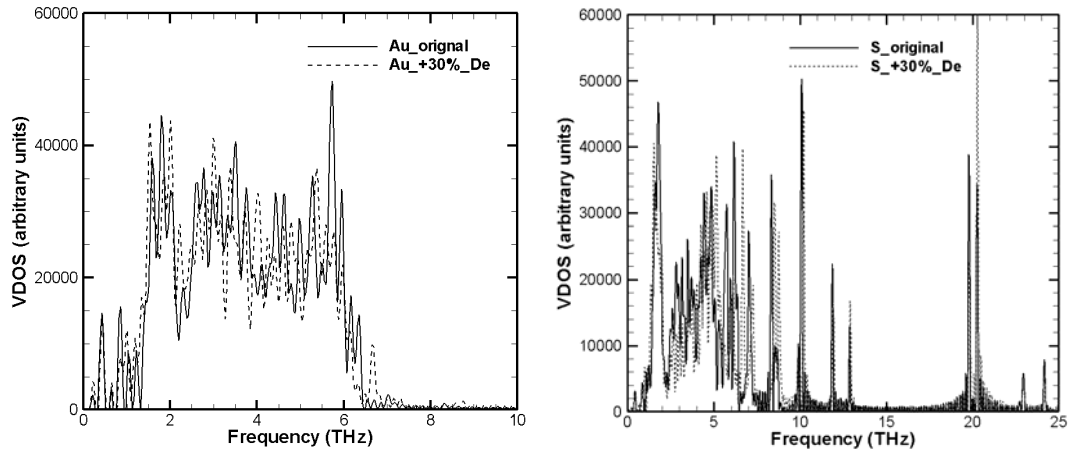

Figure 9. VDOS's of the surface Au atoms and the S atoms for the original binding energy and 130% binding energy.

**(6) Heat pulse**

To study the thermal energy dissipation in the Au-SAM-Au junctions, a heat pulse is introduced to the source region after the system reached equilibrium state at 100K. The heat pulse is imposed by scaling the temperature of the source region to 800K. To visualize the temperature profiles at different instances while minimize the influence of temperature fluctuations, time steps are divided into blocks with each block containing 1000 time steps (=500fs), temperatures are then averaged over the 1000 steps. Snapshots of temperature profiles are presented in Figure 10, and the time block in which the heat pulse was imposed was set to t=0.

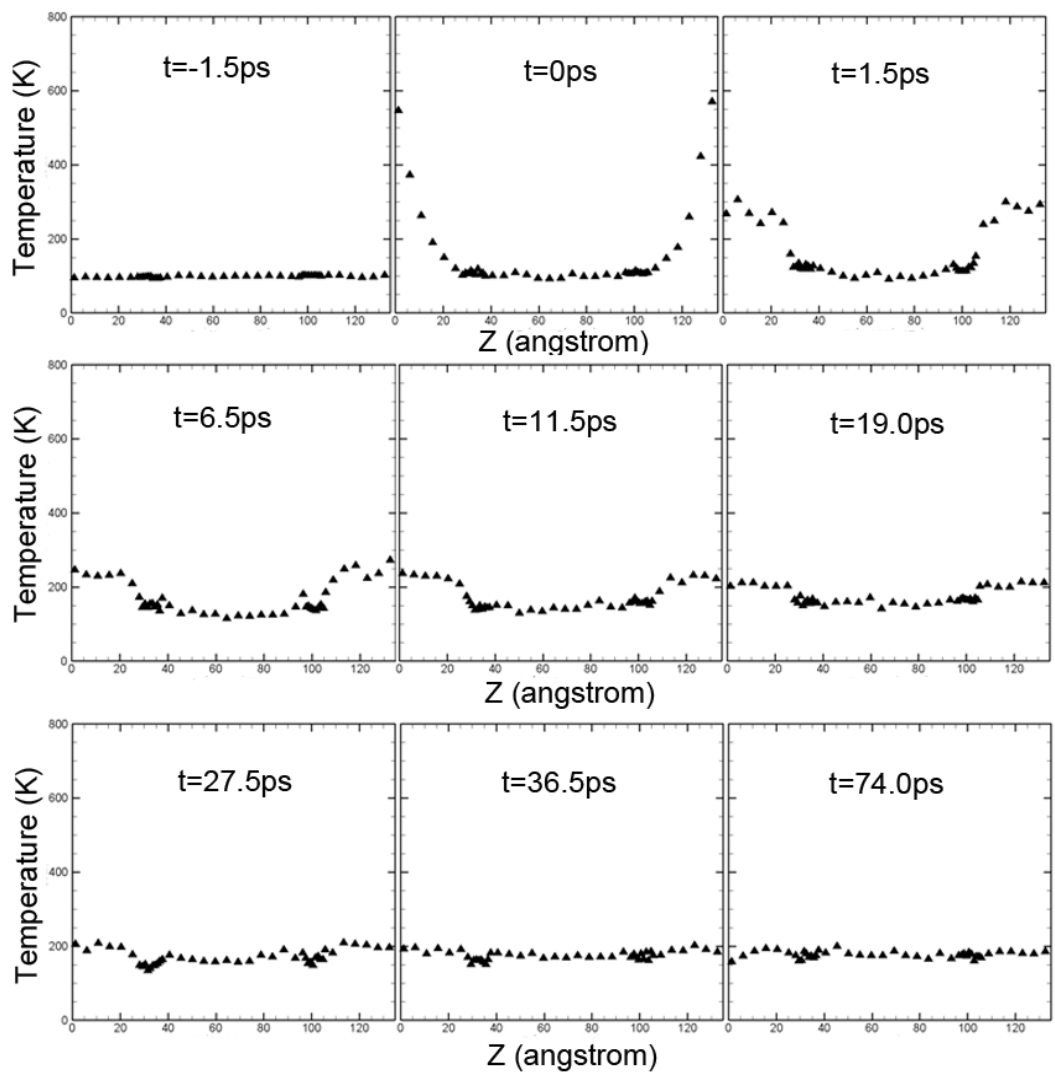

Figure 10. Snapshots of Temperature Profiles

From Figure 10, it can be seen that the time required for the heat to spread over the Au substrate is less than 1.5 ps. The time required for energy to cross the interfaces is much longer, which is between 36.5 ps and 74.0 ps. Temperature gradients inside the SAM molecules are always small in these snapshots, suggesting fast heat dissipation inside the SAM molecules. These observations prove that thermal resistances of the Au substrates and the SAM molecules are much smaller than the Au-SAM interfaces resistance.

**(6) Further VDOS Analysis**

VDOS's are also calculated for the Au atoms at the Au-SAM interface and those locate inside the substrate (Figure 11). To identify the locations of the Au atoms, the 12-layer Au substrate is divided into 6 slices with the first slice being the source region. The VDOS's of the Au atoms of the third (Au_3) and the fourth (Au_4) slices are calculated, representing the inside atoms. The VDOS's of the single layer Au atoms at the interface are calculated as the surface atoms. The VDOS's are calculated from the trajectories from an equilibrium MD run at 150K. By comparing the spectra, it can be seen that the vibrational features of the inside Au atoms are almost the same (Au_3 and Au_4), while the dynamics of the Au atoms at the interface are obviously different. Compared to the VDOS's of the inside Au atoms, the VDOS of the surface Au atoms has two obvious peaks around 2THz, and the modal populations from 3 to 6THz decreased. Overall, the VDOS moved to the low frequency side when the Au atoms are at the interface. The effective total bond strengths of the surface Au atoms are reduced due to the reduced number of bonds compared to the atoms inside the substrate. This leads to the reduced intensity of higher frequency vibrations and the emphasis of the lower frequency vibrations. Such change of spectra leads to additional boundary thermal resistance, and this is reflected in the nonlinear substrate temperature profiles near the interfaces in Figure 2.

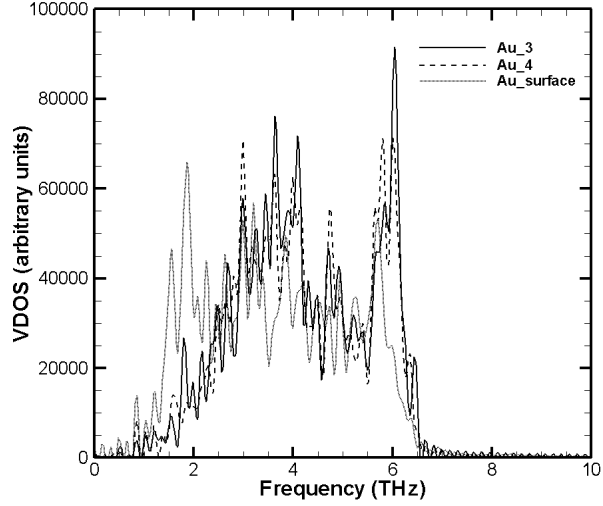

Figure 11. VDOS's of the Au atoms at different locations of the substrate

Since the areas under the VDOS lines are proportional to the temperatures, we can compare the temperatures of the LF modes and the IF modes of the SAM molecules separately. Such a frequency dependent temperature of a certain species $\alpha$ is defined in the same way as described in ref. [42]:

$$T_\alpha(\omega) = T_{\alpha,eq} \frac{\int_\omega^{\omega+\Delta\omega} D_{\alpha,ne}(\omega')d\omega'}{\int_\omega^{\omega+\Delta\omega} D_{\alpha,eq}(\omega')d\omega'} \qquad (6)$$

where $D_{\alpha,ne}(\omega)$ and $D_{\alpha,eq}(\omega)$ are the VDOS's from non-equilibrium and equilibrium runs, $T_\alpha(\omega)$ is the frequency dependent temperature of the non-equilibrium system, and $T_{\alpha,eq}$ is the temperature in an equilibrium run. The temperature of the LF modes is obtained from eq. 6 with integrals taken from 0 to 15THz, and the temperature of the IF modes is the integral from 15THz to 30THz. The non-equilibrium VDOS's are calculated from a run with a mean temperature of 100K. The temperatures are calculated for the C atoms in the SAM molecules, and the results are shown in Figure 12. The x-axes represent the numbering of the C atoms which is defined in the way as described in Section 3.2. It is very interesting to find that the temperatures of the LF modes have a symmetric profile with a convex at the center. The higher temperatures of the central C atoms suggest an extra heat source at these positions. This should be the result of the delocalized LF modes which is reported to delocalize over four to five carbon segments [6,41]. As a result, we believe some energy is transported to the central C atoms of the SAM molecules directly from the Au substrate without passing through other atoms in the molecule. Wang et. al. [6] also concluded that Au layer did not transfer its heat to an individual atom at the base of the SAM. However, we cannot confirm such a finding given the information from Figure 12. In the IF modal temperatures, the atoms that are closer to the hot substrate (solid line) have slightly higher temperatures than the atoms close to the cool substrate (dashed line). It seems that the temperature of the IF modes are weakly influenced by the substrate temperatures. Since there is no direct coupling of these modes to the substrate, the aforementioned finding is believed to be related to the different strengths of the anharmonicities of the two substrates. Overall, both temperatures of the LF modes and the IF modes are largely decoupled with the substrate temperatures. This is because that the ballistic energy transport inside a SAM molecule has a much faster rate than the interface energy transport. One could picture the thermal transport across the Au-SAM-Au junction as follow: thermal energy from the high temperature substrate is transported to the low temperature substrate through the SAM molecules (not necessarily along the molecule chains), meanwhile, energy redistributes inside the SAM at a much faster rate.

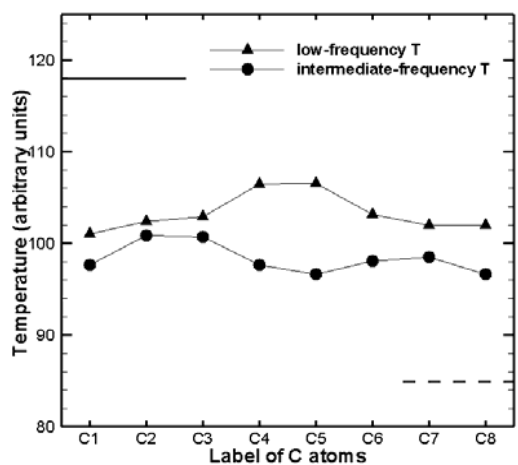

Figure 12. Frequency dependent temperatures of the LF and IF modes of the C atoms in the SAM molecules. The solid line is the surface temperature of the high temperature substrate and the dashed line is the surface temperature of the low temperature substrate.

## 4. Summary and Conclusion

The present work studied interfacial thermal conductance of Au-SAM (octanedithiol)-Au junctions using NEMD simulations. Temperature dependence, simulated pressure dependence, SAM molecule coverage dependence and bond strength dependence of the Au-SAM interfacial thermal conductance were studied. It was found that the thermal conductance increases with temperature increase at temperatures below 250K, and it did not change much at temperatures from 250K to 400K. The increase of thermal conductance was due to the increased populations of phonons which were involved in the thermal energy transport. While at high temperatures, the anharmonicity depressed the further increase of the thermal conductance. Pressure dependence was not found for the interfacial thermal conductance. Changing SAM coverage was found to be an effective way of changing the efficiency of thermal energy transport across the Au-SAM interfaces. The thermal conductance decreased obviously with coverage decrease due to the reduced number of SAM molecules, which acted as energy transport "channels". Bond strength also had an impact on the interfacial thermal transport. Stronger Au-S bonds lead to smaller anharmonicity at the interface, and thus facilitate thermal transport. VDOS's were analyzed to explore the mechanism of the Au-SAM thermal energy transport and energy transport inside the SAM molecules. The low VDOS's of the SAM molecules worked as bottlenecks, and thermal energy transport from Au to SAM molecules was not efficient. Thus the interfacial resistance was large. The VDOS's also showed that the vibration modes were largely delocalized in the SAM molecule, and the harmonic (ballistic) transport redistributed energy efficiently inside the molecules. The junction response to a heat pulse demonstrated that the heat dissipation inside the SAM molecules and inside the Au substrates were very fast, but the energy transport across the Au-SAM interface took a much longer time. These observations indicated that it was the Au-SAM interface that dominated the thermal transport across the junction. Our calculated thermal conductance agreed reasonably well with available experimental data and other theoretical predictions. All the calculated thermal conductance values were between $150$ and $450 MW/(m^2 \cdot K)$, which was inside the range of the experimentally reported thermal conductance of metal-nonmetal interfaces $8 < G < 700\ MW/(m^2 K)$ [37-39].

## Acknowledge

The authors gratefully acknowledge the support of NSF Grant Award ID 0522594 to enable this work to be performed. The authors also thank Dr. S.D. Mahanti and Dr. N. Priezjev for valuable discussions. The project is in collaboration with Professor Arun Majumdar in UC Berkeley. The calculations were performed on the high performance computers of Michigan State University.

**References:**

[1]. Y. Loo, J. W. P. Hsu, R. L. Willett, K. W. Baldwin, K. W. West and J. A. Rogers. 2002. High-resolution transfer printing on GaAs surfaces using alkane dithiol monolayers. J. Vac. Sci. Technol. B. **20**, 2853

[2]. O. S. Nakagawa, S. Ashok, C. W. Sheen, J. Martensson and D. L. Allara. 1991. GaAs interface with Octadecyl Thiol Self-Assembled Monolayer: Structural and Electronical Properties. Jap. J. App. Phy. Vol. **30**, pp. 3759-3762

[3]. A. Koike and M. Yoneya. 1996. Molecular dynamics simulations of sliding friction of Langmuir-Blodgett monolayers. J. Chem. Phys. **105**, 6060

[4]. C. L. McGuiness, A. Shaporenko, C. K. Mars, S. Uppili, M. Zharnikov and D. L. Allara. 2006. Molecular Self-Assembly at Bare Semiconductor Surfaces: Preparation and Characterization of Highly Organized Octadecanethiolate Monolayers on GaAs(001). J. Am. Chem. Soc. **128**, 5231-5243

[5]. Z. Ge, D. G. Cahill, and P. V. Braum. 2006. Thermal Conductance of Hydrophilic and Hydrophobic Interfaces. Phys. Rev. Let. **96**, 186101

[6]. Z. Wang, J. A. Carter, A. Lagutchev, Y. K. Koh, N. H. Seong, D. G. Cahill, and D. D. Dlott. 2007. Ultrafast Flash Thermal Conductance of Molecular Chains. Science. Vol. **317**. no. 5839, pp.

787 – 790

[7]. H. A. Patel, S. Garde, and P. Keblinski. 2005. Thermal Resistance of Nanoscopic Liquid-Liquid Interfaces: Dependence on Chemistry and Molecular Architecture. <u>Nano Letters.</u> **5**, 2225

[8]. Y. Yourdshahyan, H. K. Zhang, and A. M. Rappe. 2001. n-alkyl thiol head-group interactions with the Au(111) surface. <u>Phys. Rev. B.</u> **63**, 081405

[9]. H. Gronbeck, A. Curioni, and W. Andreoni. 2000. Thiols and Disulfides on the Au(111) Surface: The Headgroup-Gold Interaction. <u>J. Am. Chem. Soc.</u> **122**, 3839

[10]. C. W. Sheen, J. Shi, J. Martensson, A. N. Parikh, and D. L. Allara. 1992. A New Class of Organized Self-Assembled Monolayers: Alkane Thiols on GaAs(100). <u>J. Am. Chem. Soc.</u> **114**, 1514

[11]. W. Andersoni, A. Curioni, H. Gronbeck. 2000. Density Functional Theory Approach to Thiols and Disulfides on Gold: Au(111) Surface and Clusters. <u>International Journal of Quantum Chemistry.</u> **80**, 598

[12]. Scott Reese, Marye Anne Fox. 1998. Self-Assembled Monolayers on Gold of Thiols Incorporating Conjugated Terminal Groups. <u>J. Phys. Chem. B.</u> **102**, 9820–9824

[13]. J. Hautman and M. L. Klein. 1989. Simulation of a monolayer of alkyl thiol chains. <u>J. Chem. Phys.</u> **91**, 4994

[14]. D. G. Cahill, W. K. Ford, K. E. Goodson. 2003. Nanoscale thermal transport. <u>Journal of Applied Physics.</u> **93**, 793.

[15]. J. M. Ziman. 1960. Electrons and Phonons. Oxford University Press, New York

[16]. S. Volz, J. B. Saulnier, M. Lallemand, B. Perrin, P. Depondt and M. Mareschal. 1996. Transient Fourier-law deviation by molecular dynamics in solid argon. <u>Phys. Rev. B.</u> **54**, 340

[17]. R. H. H. Poetzsch and H. Böttger. 1994. Interplay of disorder and anharmonicity in heat conduction: Molecular dynamics study. <u>Phys. Rev. B.</u> **50**, 15757

[18]. J. P. Crocombette, G. Dumazer, and N. Q. Hoang. 2007. Molecular dynamics modeling of the thermal conductivity of irradiated SiC as a function of cascade overlap. <u>J. App. Phys.</u> **101**, 023527

[19]. A. J. H. McGaughey, M. Kaviany. 2004. Thermal conductivity decomposition and analysis using molecular dynamics simulations, Part II. Complex silica structures. <u>International Journal of Heat and Mass Transfer</u>. **47**, 1799

[20]. J. Li, L. Porter, S. Yip. 1998. Atomistic modeling of finite-temperature properties of crystalline $\beta - SiC$, II. Thermal conductivity and effects of point defects. <u>Journal of Nuclear Materials.</u> **255**, 139

[21]. T. Luo and J. R. Lloyd. 2008. Equilibrium Molecular Dynamics Study of Lattice Thermal Conductivity/Conductance of Au-SAM-Au Junctions. Under review, <u>J. Heat Transfer.</u>

[22]. P. Chantrenne and J. L. Barrat. 2004. Finite Size Effects in Determination of Thermal Conductivities: Comparing Molecular Dynamics Results With Simple Models. <u>Journal of Heat Transfer.</u> **126**, pp. 577-585

[23]. H. J. Castejon. 2003. Nonequilibrium Molecular Dynamics Calculation of the Thermal Conductivity of Solid Materials. <u>J. Phys. Chem. B.</u> 2003, **107**, pp. 826-828

[24]. C. Olischlger and J. C. Schon. 1999. Simulation of thermal conductivity and heat transport in solids. <u>Phys. Rev. B.</u> **59**, pp. 4125-4133

[25]. E. Lussetti, T. Terao, and F. M. Plathe. 2007. Nonequilibrium Molecular Dynamics Calculation of the Thermal Conductivity of Amorphous Polyamide-6,6. <u>J. Phys. Chem. B.</u> **111**, pp.

11516-11523

[26]. S. Mahajan, G. Subbarayan, and B. G. Sammakia. 2006. Thermal conductivity of amorphous silica using non-equilibrium molecular dynamics simulations. The Tenth Intersociety Conference on Thermal and Thermomechanical Phenomena in Electronics Systems. pp. 1269-1275

[27]. F. Zhang, D. J. Isbister, and D. J. Evans. 2001. Nonequilibrium Molecular Dynamics Studies of Heat Flow in One-Dimensional Systems. International Journal of Thermophysics. **22**, pp. 135-147

[28]. F. M. Plathe. 1997. A simple nonequilibrium molecular dynamics method for calculating the thermal conductivity. J. Chem. Phys. **106**, pp. 6082-6085

[29]. M. Zhang, E. Lussetti, L. E. S. de Souza, and F. M. Plathe. 2005. Thermal Conductivities of Molecular Liquids by Reverse Nonequilibrium Molecular Dynamics. J. Phys. Chem. B. **109**, pp. 15060-15067

[30]. I. H. Sung, D. E. Kim. 2005. Molecular dynamics simulation study of the nano-wear characteristics of alkanethiol self-assembled monolayers. Appl. Phys. A. **81**, 109

[31]. R. Mahaffy, R. Bhatia, and B. J. Garrison. 1997. Diffusion of a Butanethiolate Molecule on a Au(111) Surface. J. Phys. Chem. B. **101**, 771

[32]. R. C. Lincoln, K. M. Koliwad, and P. B. Ghate. 1967. Morse-Potential Evaluation of Second- and Third-Order Elastic Constants of Some Cubic Metals. The Physical Review. **157**. 463

[33]. W. L. Jorgenson. 1989. Intermolecular potential functions and Monte Carlo simulations for liquid sulfur compounds. J. Phys. Chem. **90**, 6379-6388

[34]. J.D. Gale and A.L. Rohl. 2003. The General Utility Lattice Program, Mol. Simul. **29**, 291

[35]. R. Y. Wang, R. A. Segalman, A. Majumdar. 2006. Room temperature thermal conductance of alkanedithiol self-assembled monolayers. Appl. Phys. Lett. **89**, 173113

[36]. B. C. Daly and H. J. Maris. 2002. Calculation of the thermal conductivity of superlattices by molecular dynamics simulations. Physica B. **316**, 247-249

[37]. R. J. Stoner and H. J. Maris. 1993. Kapitza conductance and heat flow between solids at temperatures from 50 to 300 K. Phys. Rev. B. **48**, 16373

[38]. R. M. Costescu, M. A. Wall, and D. G. Cahill. 2003. Thermal conductance of epitaxial interfaces. Phys. Rev. B. **67,** 054302

[39]. A. N. Smith, J. L. Hostetler, and P. M. Norris. 2000. Thermal boundary resistane measurements using a transient thermoreflectance technique. Microscale Thermophysical Engineering. **4**, 51

[40]. B. Hu, B. Li, and H. Zhao. 1998. Heat conduction in one-dimensional chains. Phys. Rev. E. 57, 2992-2995

[41]. D. Segal, A. Nitzan and P Hanggi. 2003. Thermal conductance through molecular wires. J. Chem. Phys. 119, 6840-6855

[42]. N. Shenogina, P. Keblinski, and S. Garde. 2008. Strong frequency dependence of dynamical coupling between protein and water. J. Chem. Phys. 129, 155105

[43]. M. P. Allen and D. J. Tildesley. 1987. Computer Simulation of Liquids. Oxford University Pres., Oxford.